# Efficient and Accurate Hyperspectral Pansharpening Using 3D VolumeNet and 2.5D Texture Transfer

Yinao Li, Yutaro Iwamoto, Ryousuke Nakamura, Lanfen Lin, Ruofeng Tong, Yen-Wei Chen

*Abstract*—Recently, convolutional neural networks (CNN) have obtained promising results in single-image SR for hyperspectral pansharpening. However, enhancing CNNs' representation ability with fewer parameters and a shorter prediction time is a challenging and critical task. In this paper, we propose a novel multi-spectral image fusion method using a combination of the previously proposed 3D CNN model VolumeNet and 2.5D texture transfer method using other modality high resolution (HR) images. Since a multi-spectral (MS) image consists of several bands and each band is a 2D image slice, MS images can be seen as 3D data. Thus, we use the previously proposed VolumeNet to fuse HR panchromatic (PAN) images and bicubic interpolated MS images. Because the proposed 3D VolumeNet can effectively improve the accuracy by expanding the receptive field of the model, and due to its lightweight structure, we can achieve better performance against the existing method without purchasing a large number of remote sensing images for training. In addition, VolumeNet can restore the high-frequency information lost in the HR MR image as much as possible, reducing the difficulty of feature extraction in the following step: 2.5D texture transfer. As one of the latest technologies, deep learning-based texture transfer has been demonstrated to effectively and efficiently improve the visual performance and quality evaluation indicators of image reconstruction. Different from the texture transfer processing of RGB image, we use HR PAN images as the reference images and perform texture transfer for each frequency band of MS images, which is named 2.5D texture transfer. The experimental results show that the proposed method outperforms the existing methods in terms of objective accuracy assessment, method efficiency, and visual subjective evaluation.

*Keywords—pansharpening, super-resolution, convolutional neural network*

## I. Introduction

Multispectral (MS) images, as a typical kind of remote sensing data, are widely used in agriculture, mining, and environmental monitoring applications. An ideal situation is that we can directly capture remote sensing images with high spatial resolution and high frequency spectrum. However, due to the limitation of physical constraints, we can only obtain MS images with low spatial resolution and panchromatic (PAN) images with high spatial resolution. Therefore, MS image fusion based on PAN images became a hot research topic and this technology is also called pansharpening.

Thanks to the development of deep learning technology, pansharpening research is constantly breaking through in recent years. In 2016, Masi et al. [1] proposed a new pansharpening method based on the SRCNN [2-3]. They concatenated the original PAN image and interpolated LR MS image and then use the synthetic image and HR MS image as the input and output respectively to train an SRCNN model. To improve the accuracy, Yang et al. [4] proposed a deep network architecture for pansharpening called PanNet using ResNet. By adding a high pass filter, they let the CNN model only learn the mapping relationship between the ultra-high frequency components of the MS image and PAN image and the high frequency information lost in the HR MS image. Similarly, Scarpa et al. [5] used anther deep CNN model VGG [6] instead of SRCNN to fuse LR MS image and HR PAN image. Peng et al. [7] introduced a novel pansharpening method based on a multiscale dense network (PSMD-Net) using originally designed structure. Since an MS remote sensing image generally consists of 4 or 8 frequency bands and each band can be seen as a 2D image slice, using 3D CNN can get more accurate results due to the larger receptive field compared with 2D CNN. Thus, Xie et al. [8] presented a hyperspectral pansharpening method using 3D generative adversarial networks (HPGAN). They built a deep 3D CNN model as the generator to fuse interpolated LR MS image and HR PAN image and the main feature fusion module is several 3D residual blocks. Although HPGAN outperformed previous methods in terms of accuracy, the computation cost is high. In addition, we have to prepare a large number of training samples due to a large number of parameters in HPGAN, which increases the economic cost of training the model.

Thus, in this paper, we propose a novel pansharpening processing framework using 3D VolumeNet [9] and 2.5D texture transfer to solve the above problems. Since the previously proposed 3D VolumeNet can effectively improve the accuracy by expanding the receptive field of the model, and due to its lightweight structure, we can achieve better performance against the state-of-the-art method without purchasing a large number of remote sensing images for training. In addition, VolumeNet can restore the high-frequency information lost in the HR MR image as much as possible, reducing the difficulty of feature extraction in the subsequent texture transfer. On the other hand, as one of the latest technologies, texture transfer based on deep learning is demonstrated to effectively improve the visual performance and quality evaluation indicators of image reconstruction. Different from the texture transfer processing of RGB image, we use HR PAN images as the reference images to perform texture transfer for each frequency band of MS images. The experimental results show that our method outperforms the state-of-the-art methods in terms of accuracy and the computational complexity is small.

## II. Proposed Method

As shown in Fig. 1, the proposed method for multispectral image pansharpening consists of two steps: 1) 3D deep learning model based multispectral image fusion using proposed VolumeNet; 2) 2.5D texture transfer using the SOTA 2D texture transfer CNN proposed for RGB image SR.

Let $\mathbf{X} \in \mathbb{R}^{W \times H \times L}$ be the desired SR MS image, where $W \times H$ is the number of spatial pixels and $L$ is the number of spectral channels (bands). Our goal is to reconstruct **X** by the combination of an LR MS image $\mathbf{M} \in \mathbb{R}^{(W/s)\times(H/s)\times L}$ and an HR PAN image $\mathbf{P} \in \mathbb{R}^{W\times H\times 1}$ of the same scene, where $s$ denotes the scale factor.

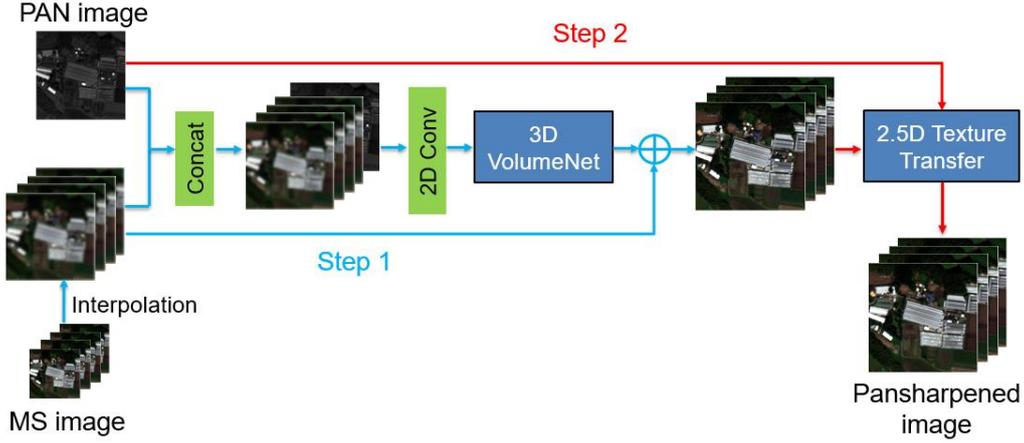

Figure 1: Framework of the proposed method for multispectral pansharpening.

In step 1, the objective is to reconstruct a preliminary $\mathbf{X}'$ using the concatenation of interpolated an SR MS image $\mathbf{M}' \in \mathbb{R}^{W\times H\times L}$ and an HR PAN image **P**. Before training the 3D VolumeNet model, we need to keep the size of the input image and output image the same. Thus, a 2D convolution layer using $1 \times 1$ kernel was built to adjust the size of concatenated $\mathbf{M}'$ and **P** to $W \times H \times L$. Then, we train a 3D VolumeNet, which is briefly described in [9], to predict the residual information of SR MS images and interpolated LR MS images from the combined graph of HR PAN images and interpolated LR MS images.

Fig. 2 illustrates the framework of Step 2 of our method. SR MS-*n* represents the *n*-th band of SR MS image processed by VolumeNet. **LR MS-*n*↑** denotes the *n*-th band of raw MS image enlarged by bicubic interpolation. **PAN** is the raw PAN image. **PAN↓↑** is the bicubic interpolated PAN image using downsampling and upsampling with the same factor $4\times$ sequentially. Thus, **PAN↓↑** is domain-consistent with **LR MS-*n*↑**. As introduced in [10], there are four parts in the texture transformer: the learnable texture extractor, the relevance embedding module, the hard-attention module for feature transfer and the soft-attention module for feature synthesis. However, the architecture in [10] was proposed for RGB image SR, while our data are hyperspectral images. Thus, we use the raw PAN image as the reference image, and perform texture transfer on each band of the MS image, which is called 2.5D texture transfer. There are four parts in the texture transformer: the learnable texture extractor, the relevance embedding module, the hard-attention module for feature transfer and the soft-attention module for feature synthesis. Details will be briefly described below.

**Learnable Texture Extractor.** In reference image-based SR tasks, texture extraction for reference images is essential because accurate and proper texture information will assist the generation of SR images. In this part, a learnable texture extractor, whose parameters will be updated during end-to-end training, is used. Such a design encourages a joint feature learning across the input LR and reference image (i.e., SR MS-*n* and PAN), in which more accurate texture features can be captured. Then, the process of texture extraction can be expressed as:

$$Q = \text{LTE}(\textbf{LR MS-}n\uparrow), \qquad (1)$$
$$K = \text{LTE}(\textbf{PAN}\downarrow\uparrow), \qquad (2)$$
$$V = \text{LTE}(\textbf{PAN}), \qquad (3)$$

where LTE( ) denotes the output of the learnable texture extractor. The extracted texture features, $Q$ (query), $K$ (key), and $V$ (value) indicate three basic elements of the attention mechanism inside a transformer.

**Relevance Embedding.** To embed the relevance between the LR and Ref image by estimating the similarity between $Q$ and $K$, I use a relevance embedding structure. At first, both Q and K are unfolded into patches, denoted as $\mathbf{q}_i$ ($i \in [1, H_{\textbf{LR MS-}n\uparrow} \times W_{\textbf{LR MS-}n\uparrow}]$) and $\mathbf{k}_j$ ($j \in [1, H_{\textbf{PAN}\downarrow\uparrow} \times W_{\textbf{PAN}\downarrow\uparrow}]$), respectively. Then, for each patch $\mathbf{q}_i$ in $Q$ and $\mathbf{k}_j$ in $K$, the relevance $r_{i,j}$ between these two patches is computed by normalized inner product:

$$r_{i,j} = \langle \frac{\mathbf{q}_i}{\|\mathbf{q}_i\|}, \frac{\mathbf{k}_j}{\|\mathbf{k}_j\|} \rangle. \qquad (4)$$

The relevance is further used to obtain the hard-attention map and the soft-attention map.

**Hard-Attention.** In this part, a hard-attention module is used to transfer the HR texture features $V$ from the reference image **PAN**. First, a hard-attention map $H$, in which the *i*-th element $h_i$ ($i \in [1, \text{HLR} \times \text{WLR}]$) is calculated from the relevance $r_{i,j}$:

$$h_i = \underset{j}{\text{argmax}}\, r_{i,j}, \qquad (5)$$

where $h_i$ is a hard index that represents the most relevant position in the reference image **PAN** to the *i*-th position in **SR MS-*n***. To obtain the transferred HR texture features $T$ from the reference image **PAN**, an index selection operation is utilized to the unfolded patches of $V$ using the hard-attention map as the index:

$$t_i = v_{h_i} \qquad (6)$$

where $t_i$ denotes the value of $T$ in the *i*-th position, which is selected from the $h_i$-th position of $V$. As a result, a HR feature representation $T$ for the input image **SR MS-*n***, which will be further used in the following soft-attention module, can be obtained.

**Soft-Attention.** In the last step of texture transfer processing, a soft-attention module is used to synthesize features from the transferred HR texture features $T$ and the LR features $F$ of the LR image. During the synthesis process, relevant texture transfer should be enhanced while the less relevant ones should be relived. To achieve that, a soft-attention map $S$ is computed from $r_{i,j}$ to represent the confidence of the transferred texture features for each position in $T$:

$$s_i = \max_j r_{i,j}, \quad (7)$$

where $s_i$ denotes the *i*-th position of the soft-attention map $S$. Then, the HR texture features T is fused with the LR features $F$ to leverage more information from the image **SR MS-*n***, and these fused features are further element-wisely multiplied by the soft-attention map $S$ and added back to $F$ to get the final output of the texture transformer, which can be represented as:

$$\mathbf{F_{out}} = F + Conv(Concat(F,T)) \odot S \quad (8)$$

where $\mathbf{F_{out}}$ indicates the synthesized output features. *Conv* and *Concat* represent a convolutional layer and concatenation operation, respectively. The operator $\odot$ denotes element-wise multiplication between feature maps.

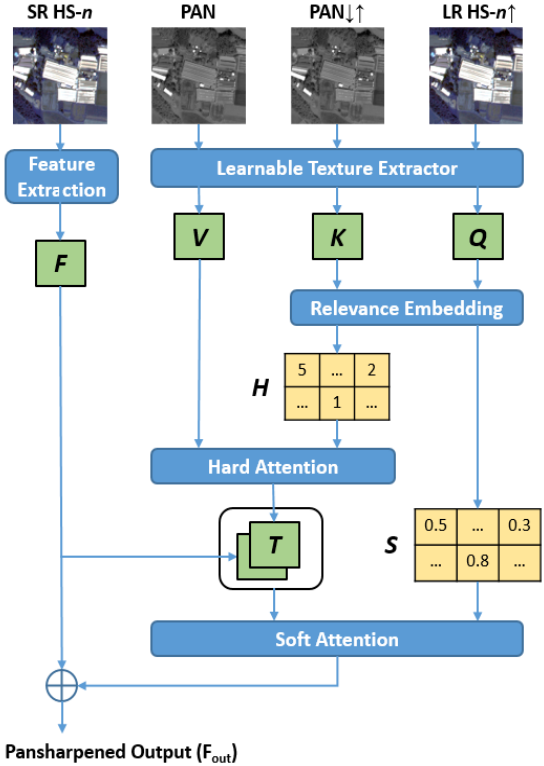

Figure 2: Framework of the second step (i.e., 2.5D texture transfer). Herein, **PAN** is the raw PAN image. **PAN↓↑** is the interpolated PAN image using downsampling and upsampling successively. **SR MS-*n*** represents the *n*-th band of SR MS image processed by VolumeNet. **LR MS-*n*↑** denotes the *n*-th band of raw MS image enlarged by bicubic interpolation.

## III. EXPERIMENTAL RESULTS

### A. Data Preparation and Training Settings

**Data of Remote Sensing Image Pansharpening.** We used WorldView3 satellite data of spatial resolution 0.31m and 8 bands, which were provided by Artificial Intelligence Research Center (AIRC), National Institute of Advanced Industrial Science and Technology (AIST), Japan. The data covered different areas, including urban, village, field, island, and seaside. We divided the MS and PAN images into 2000 patch pairs of 64 ×64 and 256 ×256 pixels, respectively. Then, we mixed the original data and randomly selected 640, 192, and 192 patch pairs for training, validation, and testing, respectively. These data were used as GND HR images and degraded to LR images at a ×4 scale in the X, Y directions via bicubic interpolation.

**Training Setting.** We implemented all models in Keras with an NVIDIA Quadro RTX 8000 GPU. The operating system and central processing unit were Ubuntu 16.04 LTS and an Intel Core i9-9820X, respectively.

We randomly extracted 32 patch pairs of size 16 ×16 ×8 (MS images) and 64 ×64 ×1 (PAN images) voxels as inputs in each training batch of the proposed 3D VolumeNet. Then we fixed the weight of 3D VolumeNet and extracted 16 patch pairs, in which MS and PAN images are of size 40 × 40 × 8 and 160 ×160 ×1 voxels, respectively, as inputs for training the texture transfer model. The other state-of-the-art models followed the settings found in the corresponding literature.

Here, we used the Adam optimizer and set the learning rate to 1e-3 and 1e-4 for training 3D VolumeNet and texture transfer model, respectively. Both of the CNNs in step 1 and 2 were optimized by the L1 norm function. We set the number of epochs to infinity, and training was terminated when the loss function had 30 subsequent epochs without reduction. Although there was no obvious overfitting according to each model's recorded convergence value during training, we only kept the weight when the model achieved the best accuracy to avoid possible overfitting. It is worth noting that the VolumeNet has to be trained and then the weights are fixed before train the texture transfer model in step 2.

### B. Results

Table 1 shows the quantitative results (i.e., mean PSNR, SSIM, CC, SAM, ERGAS, and number of parameters, and prediction time) of our proposed models, and three SOTA methods: PNN [1], PanNet [4], and PSMD-Net [7] on WorldView3 satellite image for scale factors ×4. The previously presented VolumeNet without texture transfer outperforms the SOTA methods. The proposed VolumeNet with texture transfer has the best performance among all methods and it performs much better than existing pansharpening methods.

To fully evaluate the effectiveness of our method, we present visual comparisons on scales ×4 in Fig. 3. The results indicate that the proposed method performed well in a real scenario. From the actual image processing results, it can be seen that the 2D CNN-based methods, i.e. PNN, PanNet, and PSMD-Net, ignore the connection between the various frequency bands, so the color information of MS images cannot be well integrated. Our method used 3D CNN for feature fusion; hence, it can maintain the original information of each frequency band of the original MS image very well. In addition, two steps in our proposed method can restore the spatial details and spectral details of the MS image to a great extent.

## IV. CONCLUSION

In this paper, we proposed a new hyperspectral pansharpening method using a combination of VolumeNet and 2.5D texture transfer method using other modality HR images. We used the previously proposed VolumeNet to fuse HR PAN images and bicubic interpolated MS images. Then, we used HR PAN images as the reference images and perform texture transfer for each frequency band of MS images, which is called 2.5D texture transfer. The experimental results show that the proposed method has superior performance than the existing methods in terms of objective accuracy assessment, method efficiency and visual subjective evaluation.

Table 1: Quantitative evaluation (PSNR (dB), SSIM, CC, SAM, ERGAS, and number of parameters) of different methods on WorldView3 satellite image for scale factors ×4. The arrows ↓ and ↑ under each evaluation index indicate that the ideal value is 0 and infinity, respectively

|  | PSNR (↑) | SSIM (↑) | CC (↑) | SAM (↓) | ERGAS (↓) | #param. (M) | Time (s) |
|---|---|---|---|---|---|---|---|
| PNN [1] | 30.43 | 0.8782 | 0.9752 | 0.072 | 4.627 | 0.055 | 0.358 |
| PanNet [4] | 30.03 | 0.8627 | 0.9677 | 0.075 | 5.089 | 0.079 | 0.362 |
| PSMD-Net [7] | 31.15 | 0.8640 | 0.9713 | 0.072 | 4.280 | 3.306 | 0.369 |
| VolumeNet [9] | 31.17 | 0.8881 | 0.9742 | 0.068 | 4.260 | 0.144 | 0.372 |
| VolumeNet + Texture Transfer | **31.75** | **0.8886** | **0.9747** | **0.067** | **4.040** | 1.126 | 0.534 |

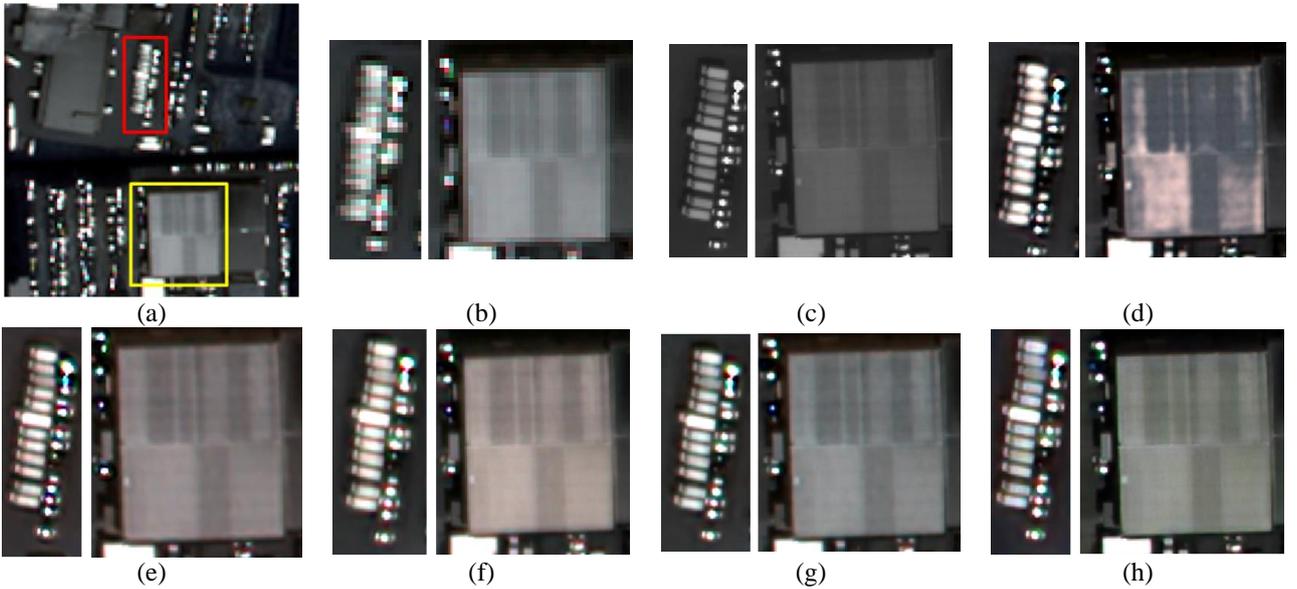

Figure 3: Qualitative results of the proposed method and typical state-of-the-art methods. (a) WorldView3 satellite image sample, (b) HS image interpolated by tricubic, (c) PAN image, (d) result of PNN, (e) result of PanNet, (f) result of PSMD-Net, (g) result of VolumeNet, (h) result of VolumeNet with texture transfer.